\documentclass{PoS}

\usepackage[fleqn]{amsmath}

\def\be{\begin{equation}}
\def\ee{\end{equation}}
\def\bga{\begin{gather}}
\def\ega{\end{gather}}
\newcommand{\baln}{\begin{align}}
\newcommand{\ealn}{\end{align}}
\newcommand{\myendalign}{\end{align}}
\def\bea{\begin{eqnarray}}
\def\eea{\end{eqnarray}}

\def\bal#1\eal{\begin{align}#1\end{align}} 

\def\lsim{\mathrel{\rlap{\lower4pt\hbox{\hskip1pt$\sim$}}
    \raise1pt\hbox{$<$}}}                

\def\bgat#1\egat{\begin{gather}#1\end{gather}}

\def\calO{{\cal O}}

\newcommand{\vev}[1]{\langle #1 \rangle}

\title{Dynamical QCD simulation with $\theta$ terms}

\ShortTitle{Dynamical QCD simulation with $\theta$ terms}
\author{\speaker{Taku Izubuchi}\\
        RIKEN-BNL Research Center, Brookhaven National Laboratory, Upton, NY 11973, USA \\
Institute for Theoretical Physics, Kanazawa University, Kanazawa 920-1192, Japan\\
}
\author{Sinya Aoki\\
        RIKEN-BNL Research Center, Brookhaven National Laboratory, Upton, NY 11973, USA \\
        Graduate School of Pure and Applied Sciences, University of Tsukuba, 
        Tsukuba, Ibaraki, 305-8571, Japan}
\author{Koichi Hashimoto\\
  Institute for Theoretical Physics, Kanazawa University, Kanazawa 920-1192, Japan\\
  Radiation Laboratory, RIKEN, Wako 351-0158, Japan}
\author{Yoshifumi Nakamura\\
        John von Neumann Institute NIC/DESY Zeuthen, 15738 Zeuthen, Germany}
\author{Toru Sekido\\
  Institute for Theoretical Physics, Kanazawa University, Kanazawa 920-1192, Japan\\
  Radiation Laboratory, RIKEN, Wako 351-0158, Japan}
\author{Gerrit Schierholz\\
        Deutsches Elektronen-Synchrotron DESY, 22603 Hamburg, Germany\\
        John von Neumann Institute NIC/DESY Zeuthen, 15738 Zeuthen, Germany}
\abstract{The $\theta$ term that breaks the Strong CP symmetry
        is introduced in the two flavors of dynamical QCD simulation.
        $\theta$ is analytically continued to a pure imaginary number
        to make the  probability of Monte Carlo positive.
        The Neutron's Electric Dipole Moment (NEDM) is measured on the
        ensemble under a uniform and week electric field.  Other
        applications of $\theta$ terms are also discussed.  }

\FullConference{The XXV International Symposium on Lattice Field Theory\\
		 July 30 - August 4 2007\\
		 Regensburg, Germany}

\begin{document}
\section{Introductions}
\label{sec:introduction}
Since the first attempt to get the non-perturbative prediction of 
the Neutron's Electric Dipole Moment (NEDM) with 
the Strong CP violating $\theta$ term,
\bal
S_\theta = 
i {\theta \over 32\pi^2} 
\int \epsilon_{\mu\nu\tau\rho} 
F_{\mu\nu} F_{\tau\rho} d^4 x
= i\theta Q_\text{top}~~,
\label{eq:theta term}
\eal
on lattice 18 years ago \cite{Aoki:1989rx}, 
there have been  renewed interests on this calculation  recently 
corresponding to the new experiments and theoretical developments.
The NEDM is very sensitive to the sea quark mass, as we will see,
and the calculation could be much improved thanks to the recent 
various developments for the simulation with dynamical quarks.

All the calculations aiming for NEDM so far introduce the effect 
of $S_\theta$ in terms of reweighting technique: the statistical ensembles of QCD vacuum were
generated via the Boltzmann weight of $\theta=0$ case,
$\text{Prob}( U ) \propto \det^{N_F} D[U] e^{-S_0[U]}~~, $
and then the effect of  $S_\theta$ is treated as a part of observable 
in the measurement stage,
$\vev{\calO}_\theta = \sum_{U} \calO e^{-S_\theta} \text{Prob}( U )/
\sum_{U} 1 e^{-S_\theta} \text{Prob}( U )$.

In this study, we introduce the CP violation according to $S_\theta$ 
in the probability of QCD ensemble generation, then measure observables
without reweighting technique to see whether the new way of calculation
has better control over the statistical and the systematic errors.
To keep the Boltzmann weight positive semi definite, the value of $\theta$
is analytically continued to pure imaginary,
\begin{equation}
\theta \to -i \theta.
\label{eq:analytic continuation}
\end{equation}
Such analytic continuation was previously explored  
for CP$^{n-1}$ models with $\theta$ term \cite{Bhanot:1984rx,Imachi:2006gq} 
and QCD with the chemical potential \cite{deForcrand:2002ci}.

To measure NEDM, a constant electric field, $\vec{E}$, is applied to Nucleon, 
and the energy, $m_N$, dependence on its spin polarization, $\vec{S}$, 
is measured,
\begin{equation}
m_N( \vec{E}, \vec{S} )-m_N( \vec{E}, -\vec{S} )
= 2 i d_N(\theta) \vec{S}\cdot\vec{E} + {\cal O}(E^3)~~.
\label{eq:energy difference}
\end{equation}

We use electric field constituted from the ordinary U(1) gauge field,  
\begin{equation}
U^\text{EM}_\mu(x) = e^{i q A^\text{EM}_\mu(x)},~~ q A^\text{EM}_\mu(x)\in 
\mathbb{Re}~~,
\end{equation}{
on {\it Euclidean} lattice, oppose to its {\it Minkowski} space 
version used in the original calculation \cite{Aoki:1989rx}, 
which is reported to be susceptible to the systematic error induced 
at the periodic boundary in temporal direction \cite{Shintani:2006xr}.
To avoid the boundary effect and also to safely neglect 
the $\calO(E^3)$ in (\ref{eq:energy difference}) at a time,
a {\it uniform and weak electric field} invented some times ago 
\cite{Heller:1987ws,PhysRev.133.A1038} is implemented.

By assuming $\theta$ is small, the NEDM , $d_N(\theta)$,
is expanded in terms of $\theta$,
\begin{equation}
d_N(\theta) = d^{(1)}_N \theta + {\cal O}(\theta^3)~~,
\label{eq:dN expansion}
\end{equation}
within which approximation, the analytic continuation, (\ref{eq:analytic continuation}),
can be trivially reverted, and more importantly, 
the energy splitting (\ref{eq:energy difference}) becomes
a real number, which could be easily measured from 
the ratio between polarized propagators  of Neutron whose spin are 
parallel and anti-parallel to the direction of the electric field,
\begin{equation}
R(t,t_{src};E) =
\lim_{t\to\text{large}}
{ \vev{N_\uparrow(t) \overline N_\uparrow(t_{src})}_{\theta,E}
  \over
  \vev{N_\downarrow(t) \overline N_\downarrow(t_{src})}_{\theta,E}
}
\propto
e^{+d_N(\theta) E t}~~.
\label{eq:prop ratio naive}
\end{equation}

In the following sections, the NEDM calculation and preliminary
results are mainly detailed.  We also briefly report another related
attempt, restricting the imaginary $\theta$ action to a time slice and
identifying it as the source term for the flavor-singlet pseudoscalar
meson ($\eta'$) for disconnected quark loop measurements.

\section{Ingredients for NEDM}
\label{sec:ingredients}
Before describing the novel simulation setups,
we compare the external electric field method
employed in this work with another method,
the form factor 
calculation \cite{Shintani:2005xg, Berruto:2005hg}, in which
NEDM is calculated from the  CP-odd form factor, $F_3(q^2)$.
While the latter method is free from the $\calO(E^3)$ 
contamination in (\ref{eq:energy difference}), one needs to insert
momentum $q$ to the vector current of the Nucleon's three point 
function, and extrapolate $q^2\to 0$ limit.
We chose the simpler method, the external electric field method, 
that involves only two point functions with zero momentum, expecting 
the statistical and/or systematic error could be smaller than the other.

\subsection{Dynamical simulation with $\theta$ terms}
All the previous lattice QCD calculations with
the strong CP breaking term $S_\theta$, (\ref{eq:theta term}),
were carried out with the reweighting technique as discussed
in Sec. \ref{sec:introduction}.
Alternatively, one could put $S_\theta$ into the ensemble
probability if $S_\theta$ becomes real number by
$\theta$ being analytically continued to pure imaginary,
(\ref{eq:analytic continuation}).
There are two options in the implementation of $S_\theta$
on lattice: one is the gluonic definition from $\sum_x F\tilde F(x)$ 
with an appropriate smeared/cooled link field 
so that the resulting topological charge, $Q_\text{top}$, become
(close to) an integer value.
This implantation may be useful for a certain application
such as precise determination of the $Q_\text{top}$ 
distribution in pure Yang-Mills theory \cite{Giusti:2007tu}.
Actual simulation coding, however, might likely involve
the smeared/cooled gauge force term, and results could strongly
depend on the details of the smearing/cooling used.
Simpler implementation would be using the anomalous axial transformation
in the continuum theory,
\begin{gather}
\psi \to e^{-i\theta \gamma_5/(2N_f)} \psi,
~~\overline\psi \to \overline\psi e^{-i\theta \gamma_5/(2N_f)}\\
{\cal D}\psi{\cal D}\overline\psi \to 
{\cal D}\psi{\cal D} \overline\psi\,\,
e^{S_\theta}~~,
\label{eq:continuum anomalous rotation}
\end{gather}
that cancels the $S_\theta$ from the original action
at the cost of CP violating $\gamma_5$ mass term
introduced in  the fermion's Lagrangian  for $N_F=2$,
\begin{equation}
{\cal L}_f = \sum_{i=u,d} \overline\psi_i( \not\!\!\,D +m 
+i m {\theta\over 2}  \gamma_5)\psi_i~~
+\calO(\theta^2).
\label{eq:fermion lagrangean}
\end{equation}

We chose  the latter implementation because of 
the absence the ambiguity from cooling/smearing. 
By the analytic continuation, (\ref{eq:analytic continuation}),
the fermion determinant with the $\gamma_5$ mass term 
in (\ref{eq:fermion lagrangean}) remains positive:
$
\det( \not\!\!\,D +m +i m{\theta\over 2}\gamma_5)^2
\to 
|\det ( \not\!\!\,D +m + m{\theta\over 2}\gamma_5)|^2.
$

Our lattice fermion action is the RG improved 
clover fermion identifying the continuum counter parts
as follows:
\begin{equation}
\not\!\!\,D+m \to D^\text{clover},\\
im{\theta\over 2}\gamma_5 \to 
i(m-m^{c}) {\theta\over 2}\gamma_5~~,
\end{equation}
where $m^{c}$ is the mass point where the
pseudoscalar meson becomes massless.

We note the relation, (\ref{eq:continuum anomalous rotation})
(\ref{eq:fermion lagrangean}), are not be precisely realized 
for the clover fermion and there will be a discretization 
error which will vanish in the continuum limit. 
More demanding simulation with overlap fermion 
or domain-wall fermions might be free from such systematic error.

\subsection{Uniform and weak electric field}
\label{subsec:uniform and weak electric field}
The second ingredients for NEDM is the external electric 
field on the periodic lattice. The electric field needs to be weak and 
constant everywhere including the periodic boundaries
as discussed in Sec. \ref{sec:introduction}.
Naively descretizing constant electric field of the continuum theory, 
for example, in $z$ direction, 
$A^\text{EM}_z(x)= E t$ in terms of  vector potential, 
one obtains $U^\text{EM}_z =\exp(i q E t)$ for charge $q$ fermion field.
This Abelian link field gives constant electric field, $e^{iF_{tz}} =e^{iqE}$, everywhere
except at the boundary between $t=L_t-1$ and $t=0$, where the strong 
opposite electric field, $\exp(-iq E(L_t-1))$, is induced.
This delta function like field causes a skewed Neutron propagator
near the boundary \cite{Shintani:2006xr}. 
By using the constructions in \cite{Heller:1987ws,PhysRev.133.A1038}, such systematic error could 
be avoided by adjusting the temporal link, only at the boundary, 
as  
\begin{gather}
U^\text{EM}_t(z,L_t-1)=\exp(-i q E (L_t-1) z)~~,~~~~
E={q 2 \pi \over L_t L_z}  \times
\mathbb{Z}
\label{eq:electric field discretization}
~~,
\end{gather}
where the condition (\ref{eq:electric field discretization})
is required not to induce another opposite field 
at the corner plaquette, including points $(z,t)=(L_z-1,L_t-1)$ and $(0,0)$,
of the periodic boundaries.
We emphases this construction has weaker electric field,
$E\sim 0.01  q \times \mathbb{Z}$ than 
the simpler construction with $E=2\pi/L_t$ 
so that $\calO(E^3)$ contamination 
in (\ref{eq:energy difference}) is less problematic.

\section{Progress report on NEDM calculation}

We now report the preliminary results of NEDM calculation with the
setups described in Sec. \ref{sec:ingredients} employed on $16^3\times
32$ lattice.  The gauge action of our choice is the RG improved gauge
action a la Iwasaki with $\beta=2.1 (a^{-1}\sim 2 \text{GeV})$.
Thermalized statistical ensembles of two flavors of dynamical clover fermion
action ($c_{sw}=1.47$) with $\theta=0.4$ and $0.0$ are accumulated for
$\tau=2,000$ in the molecular dynamics time unit.  
The parameter point and the lattice scale
were determined using the mean field improvement by 
CP-PACS \cite{AliKhan:2001tx}.
The effect of
$\theta$ vanishes at massless limit as seen in (\ref{eq:fermion
  lagrangean}). As this is a feasibility study, we consciously set the
quark mass unphysically large, $\kappa=0.1357$ or $m_\pi/m_\rho \sim
0.85$. The HMC code with accelerations 
a la Sexton-Weingarten, and also Hasenbush is used \cite{Gockeler:2007rm}.

By switching on $e^{-S_\theta}=e^{-\theta Q_\text{top}}$ in the
Boltzmann weight, the negative topological charge should appear more
frequently than the positive charge for $\theta>0$.  Figure
\ref{fig:topology distribution} clearly shows such reaction to
$\theta$.  The positive and negative charge equally emerge for
$\theta=0$ in the Gaussian-like distribution, which is shifted in
negative direction, $\vev{Q}=-7.9$, for $\theta=0.4$ ensemble.  Here
we plot the distribution of the ${\cal O}(a^2)$ improved topological
charge defined from the gluon configuration,
$Q_\text{top} = {5\over 3} Q^{1\times 1} - {2\over 3} Q^{2\times 1}~~,
$
where $Q^{1\times 1}$ and $Q^{1\times 2}$ are 
topological charges  calculated  
from $1\times 1$ and $1\times 2$ clover leafs 
respectively \cite{AliKhan:2001ym}. The lattice is smoothed until 
$Q_\text{top}$ becomes saturated close to an integer value. 
Applying 200 steps of APE smearing ($c_\text{APE}=0.45$) 
turns out to be sufficient.

\begin{figure}
\includegraphics[width=0.45\textwidth]{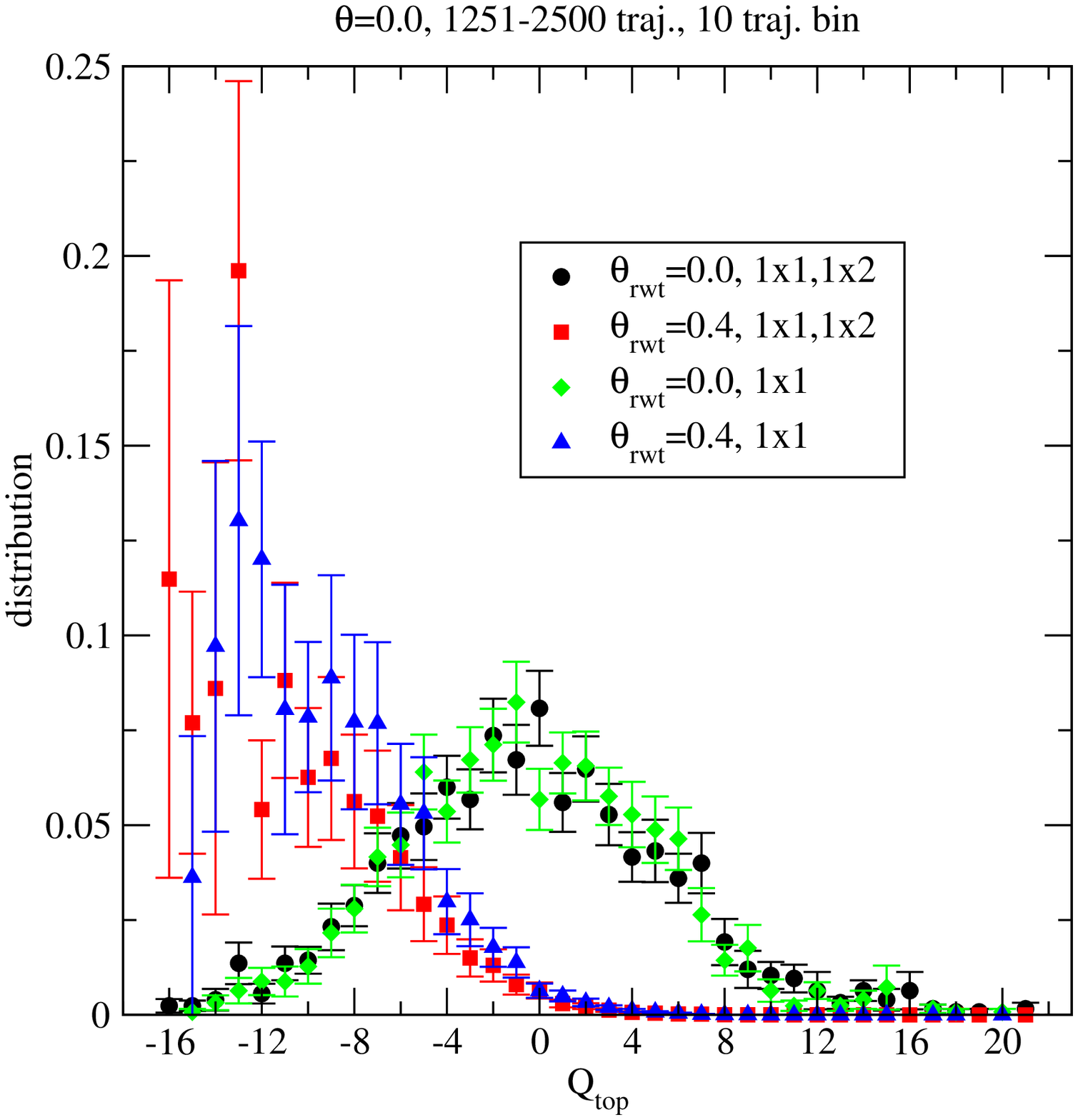}
\includegraphics[width=0.45\textwidth]{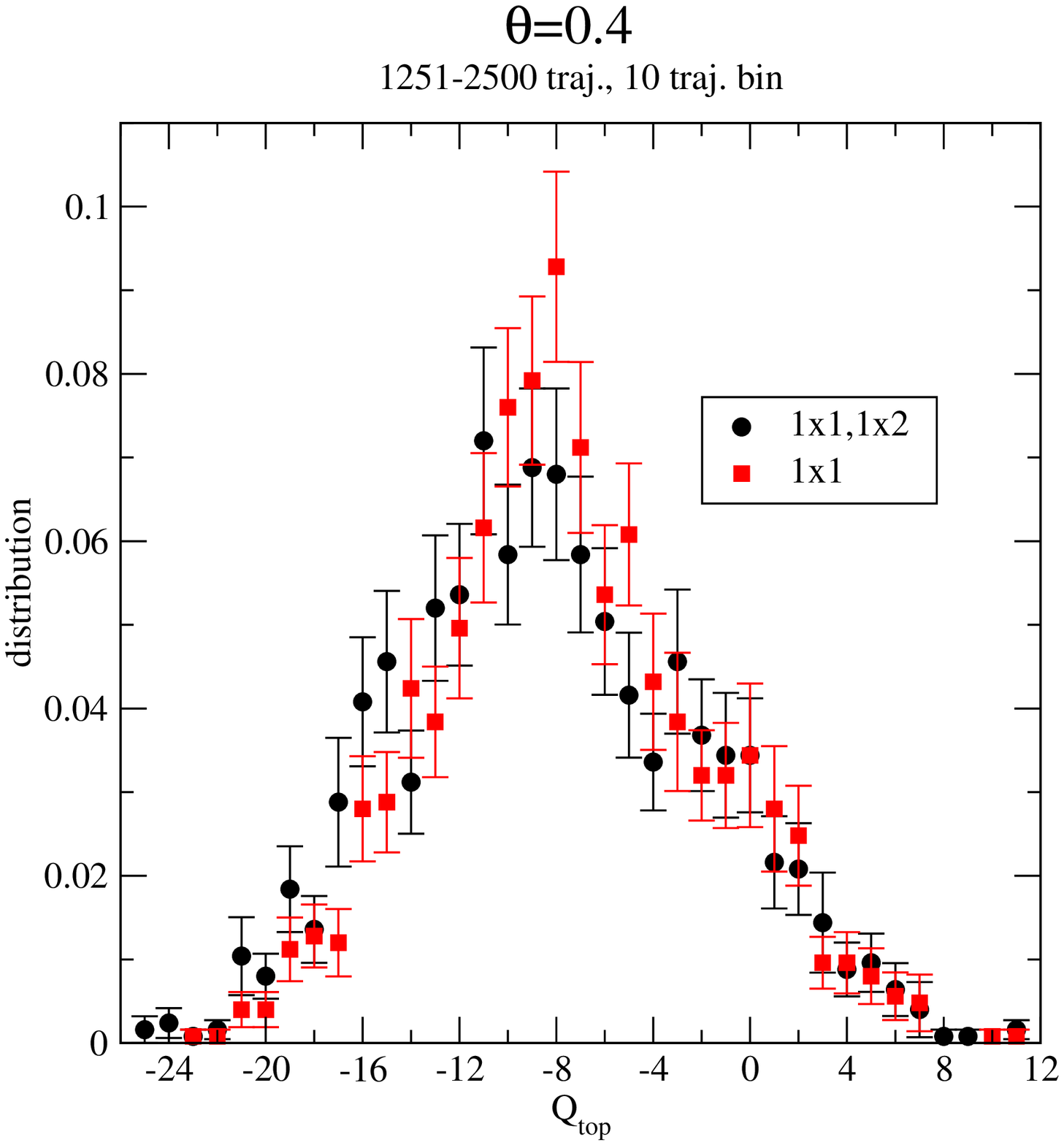}
\caption{
The distributions of $Q_\text{top}$ for $\theta=0.0$ (left) and
$\theta=0.4$ (right). $Q_\text{top}$ is calculated 
by ${\cal O}(a^2)$ improved definition on 200 steps 
of APE smeared ($c_\text{APE}=0.45$) lattices.
Also the definition only uses $1\times 1$ plaquette is plotted.
The distribution for $\theta=0.4$ by the reweighting technique 
is plotted on the left figure to show the poor predictability 
for such sizable $\theta$.
}
\label{fig:topology distribution}
\end{figure}

The polarized Neutron propagator is measured using the code
developed in \cite{Blum:2007cy} on 100 configurations 
under the external electric field, that is a multiple of,
$E_0=2\pi/(L_z L_t)\times 3\simeq 0.0368$, in $z$ direction.
The quark charges are set as $q_\text{up}=2/3$ and $q_\text{down}=-1/3$.
On each lattice, two sources for Neutron made of 
the Gaussian smeared quarks are prepared, 
and the propagators of Nucleon are measured for each sources separately. 

The spin dependent ratio of the propagator,
(\ref{eq:prop ratio naive}), is estimated from the propagator averaged
with the one with the other source points, and also 
with the one applied the opposite electric field with 
the appropriate combination as follows:
\begin{equation}
R_2(t-t_{src})=\prod_{t_{src}} \left( {R(t,t_{src}; E) \over R(t,t_{src};-E)}\right)^{1/(2 N_{src})}~~.
\end{equation}
Thanks to the temporal translational invariance of the external electric field 
described in SubSec.~\ref{subsec:uniform and weak electric field},
the backward propagator with the proper parity projection, 
which is the anti-Neutron, at $t-t_{src}<0$ may also be used 
in the average to enhance the signal. 
While a reasonable signal for $|t-t_{src}| \lsim 4$ can be seen 
within current statistics,
we need more ensemble and measurements to check the excited state
contamination. We also compare $R_2(t)$ using the traditional 
reweighted technique on $\theta=0$ lattice with same statistics, and
found about a factor of two smaller statistical error for the new method.

\begin{figure}
\hspace*{0.7cm}
\includegraphics[width=0.4\textwidth,clip]{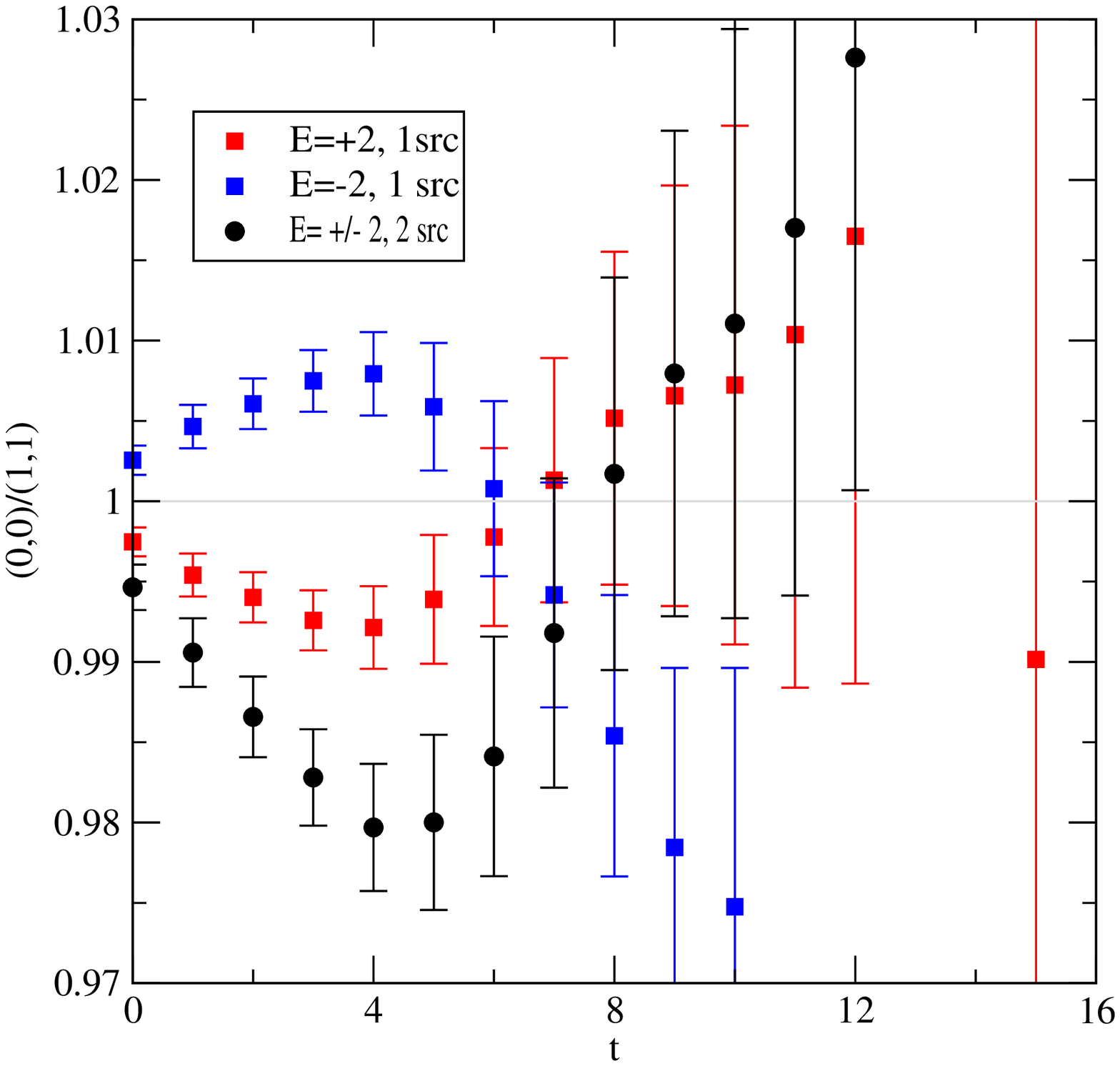}
\ \ \ \ \hspace*{1.0cm}
\includegraphics[width=0.4\textwidth,clip]{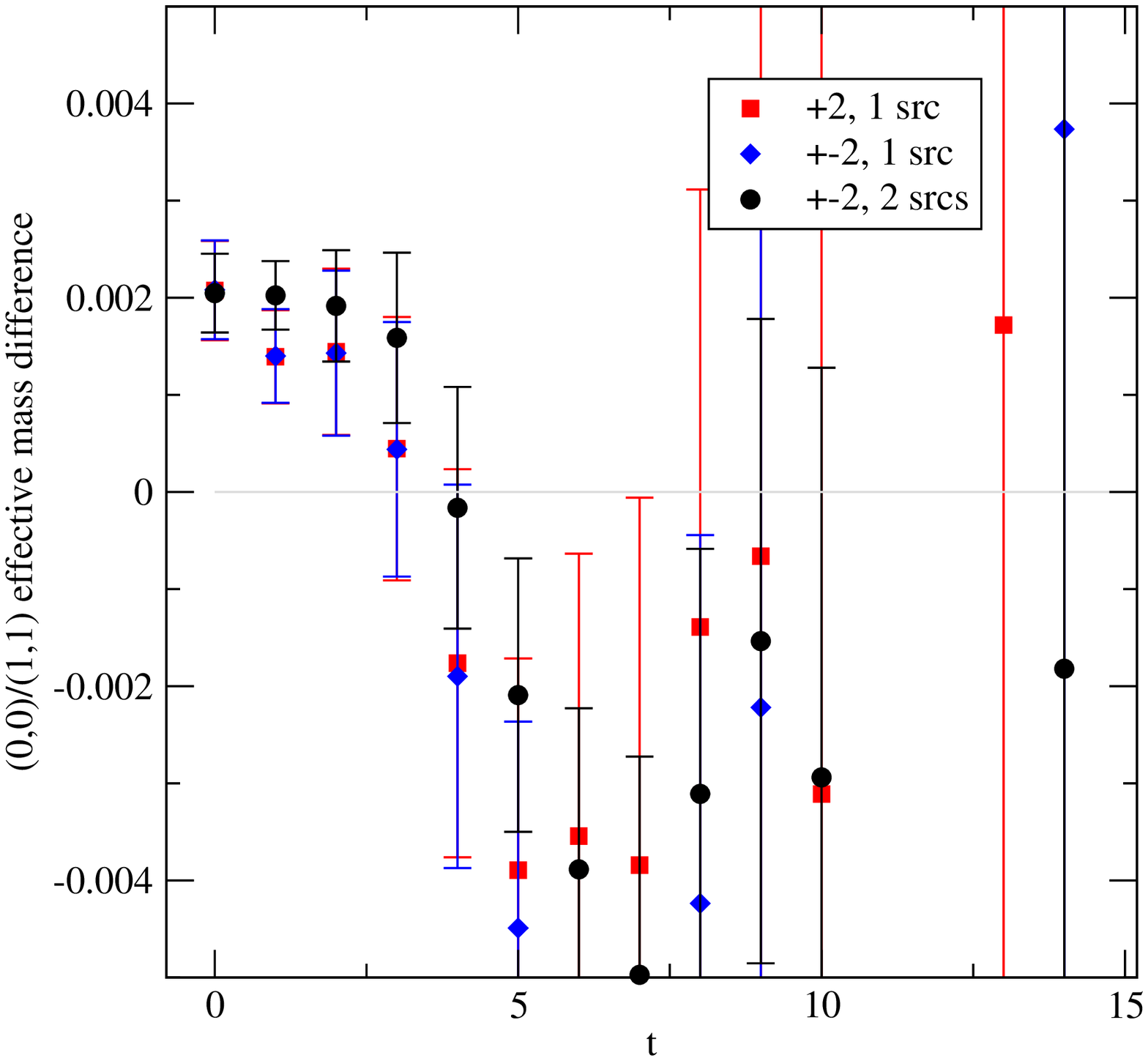}
\caption{
The double ratio, $R_2(t-t_{src})=\prod_{t_{src}}$ (left), 
and it's effective mass $\log\Big(R(t)/R(t+1)\Big)$ (right) are plotted.
The applied external electric fields are 
$E=+2E_0$ (red),
$E=-2E_0$ (blue), and the appropriate average over
$E=\pm 2E_0$ (black).
The red and green symbols are calculated using one source points
while the block symbols is the average of two source results. 
}
\label{fig:R2}
\end{figure}

Due to  lack of the chiral symmetry in clover fermion action
and also from $\calO(\theta^2)$ in (\ref{eq:fermion lagrangean}),
$\theta$ in (\ref{eq:fermion lagrangean}) may be 
mixed with different operators. To check the effect of $\theta$ in other
quantities than NEDM, we calculate the shift of topological charge 
$\vev{Q_\text{top}}_\theta$ and pseudoscalar meson mass $M_{PS}(\theta)$ 
on $\theta=0.4$ configuration.
For  crude estimates of effective $\theta$ value on this ensemble, 
we assume Gaussian distribution for $Q_\text{top}$,
and the leading order prediction of chiral perturbation theory,
$M^2_{PS}(\theta) = M^2_{PS}(0) \cosh(\theta/N_f)$ 
in \cite{Brower:2003yx}. 
The ratio of the effective $\theta$s obtained by these method 
to the input value, 0.4, turn out to be 0.6(1) from 
$\vev{Q_\text{top}}_\theta$, and 0.9(5) from pseudoscalar mass.

\section{Application for $\eta'$ calculation}
If we generate ensemble replacing $S_\theta$ by
\begin{equation}
S_\lambda = \lambda \sum_{\vec{x}} \overline \psi \gamma_5\psi(\vec{x},t_s)~~,
\end{equation}
which is the source term enhancing the negative topological charge density 
only at a time slice $t=t_s$, the resulting configurations could 
be used to measure quantities involve the disconnected quark loop.
Here we present a calculation of the flavor-singlet pseudoscalar 
meson ($\eta'$) propagator as an example.

The pseudoscalar density at 
$t$ with the Dirac operator including the $\lambda$ term, $D_\lambda$
is estimated stochastically,
\def\Tr{\text{Tr}}
\begin{equation}
\vev{\Tr  \gamma_5 D^{-1}_\lambda(t,t)}_\lambda
= - \lambda\vev{\Tr \gamma_5 D^{-1}(t,t_s) \gamma_5 D^{-1}(t_s,t)}_0
  + 2\lambda\vev{\Tr \gamma_5 D^{-1}(t,t)  \Tr \gamma_5 D^{-1}(t_s,t_s)}_0
+{\cal O}(\lambda^3)~~,
\end{equation}
whose left hand side is the $\eta'$ correlation function between
$t_s$ and $t$ omitting ${\cal O}(\lambda^3)$.
The result of effective mass on 1,000 configuration of 
$8^3\times16, a\sim 0.2 \text{fm}$ lattice is shown 
in Figure \ref{fig:eta} compared with the conventional 
two point function method (black circle).
The central value is consistent with each other, however,
the source method has much rapid error growth by increasing
$t-t_s$.

\begin{figure}
\includegraphics[width=0.4\textwidth,clip]{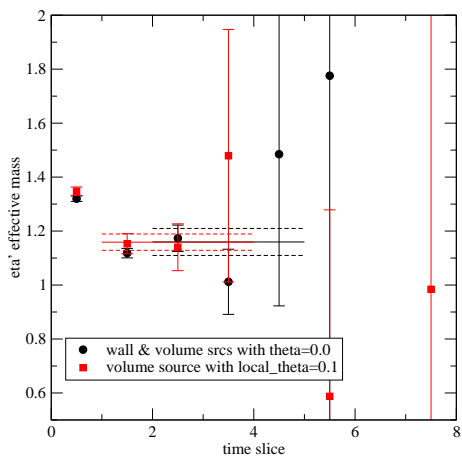}
\caption{The effective mass of the flavor singlet pseudoscalar meson
as a function of time using the source method compared with
the conventional two point function method.
}
\label{fig:eta}
\end{figure}

\section{Summary and Conclusion}
We have generated $N_F=2$ lattice QCD ensembles with the explicit
CP violation term, $\theta$ term. To preserve the positivity
of the generating probability, $\theta$ is analytically continued
to pure imaginary. By applying the uniform and weak electric field 
on the ensemble, Nucleon's electric dipole moment is calculated.
While the results are only on two parameter points,
the signal to noise ratio observed is encouraging considering 
the relatively small statistics. 
The checks for the excited state contamination by increasing
statistics, effects of the terms of higher orders  in 
$\theta$ and $E$, and the quark mass dependence are 
subjects for future study.

We also explore QCD ensembles with the source
for the flavor singlet pseudoscalar density.
The propagator of the flavor singlet pseudoscalar meson, $\eta'$, 
is calculated by restricting the imaginary $\theta$ term 
on a time slice, consistent results with the conventional 
method are obtained though the statistical error turns out
to be larger than that from the conventional method.

\section*{Acknowledgements}
We are grateful to authors and maintainers of the DESY code and CPS++,  
which are used for ensemble generation and measurements in this work.
We thank for computational resources supported by the Large Scale 
Simulation Program No. 07-14 (FY2007) of High Energy Accelerator 
Research Organization (KEK),  the RIKEN Super Combined Cluster (RSCC), 
 and QCDOC at RBRC and  Columbia University.
T.I. thanks to the authors of \cite{Blum:2007cy} for valuable discussions.



\end{document}